\documentclass[epj,final]{svjour}

\usepackage{amsmath}
\usepackage{amssymb}

\usepackage{graphicx}
\usepackage{dcolumn}
\usepackage{bm}

\newcommand{\openone}{\mathbb{I}}
\newcommand{\G}{G}

\newcommand{\eps}{\varepsilon}

\newcommand{\w}{\omega}

\newcommand{\bra}{\rangle}
\newcommand{\ket}{\langle}

\newcommand{\ex}{\textrm{exp}~}

\newlength{\figwidth}

\begin{document}

\title{Localization properties of driven disordered one-dimensional systems}

\author{Dario~F.~Martinez
\and
Rafael~ A.~Molina}

\authorrunning{D.F.\ Martinez and R.A.\ Molina}

\institute{Max-Planck-Institute f\"ur Physik Komplexer Systeme,
N\"othnitzer Str. 38, 01187 Dresden, Germany}

\date{\today}

\abstract{ \PACS{ {72.15.Rn}{Localization effects} \and
{73.20.Fz}{Weak or Anderson localization} \and {73.21.Hb}{Quantum
wires} } We generalize the definition of localization length to
disordered systems driven by a time-periodic potential using a
Floquet-Green function formalism. We study its dependence on the
amplitude and frequency of the driving field in a one-dimensional
tight-binding model with different amounts of disorder in the
lattice. As compared to the autonomous system, the localization
length for the driven system can increase or decrease depending on
the frequency of the driving. We investigate the dependence of the
localization length with the particle's energy and prove that it
is always periodic. Its maximum is not necessarily at the band
center as in the non-driven case. We study the adiabatic limit by
introducing a phenomenological inelastic scattering rate which
limits the delocalizing effect of low-frequency fields.}

\maketitle


\section{Introduction}

The seminal paper of Anderson on the absence of diffusion in
certain random lattices \cite{Anderson58} started a new era in the
study of the effects of disorder in condensed matter physics. Real
materials always contain a certain degree of disorder, since the
atomic structure is never perfectly regular, and many physical
properties are either influenced or even mostly determined by this
randomness \cite{Lifshits}. As the translational symmetry is
broken, electrons in disordered lattices are not delocalized Bloch
waves and can be spatially confined. One of the main quantities of
interest in disordered systems is the localization length
$\lambda$ of the electron wave-functions. A system of typical size
$L>\lambda$ will behave as an insulator while a system with
$L<\lambda$ will behave as a conductor \cite{Lifshits,Kramer}.

The Anderson model in one dimension has been used as a fundamental
tool for the understanding of the properties of quantum and
molecular wires since the beginning of the field of molecular
electronics \cite{Mujica94}. In one dimension it can be shown in a
mathematically rigorous way that all eigenstates of the Anderson
model are exponentially localized in the thermodynamic
limit\cite{Ishi}.

On the other hand, the use of a time-periodic driving field in
nano-technological applications has emerged in recent years as a
way to control the properties of possible future devices
\cite{Gert.molecules} and as a source of many interesting new
effects \cite{PAT}. For example, in semiconductor superlattices in
the presence of THz radiation, absolute negative conductance and
dynamical localization have been observed experimentally
\cite{Keay95,Winnerl97}. In the field of molecular electronics, a
ratchet-like effect was shown experimentally in the photocurrent
in a self-assembled monolayer of asymmetric molecules
\cite{Yasutoni2004}. In terms of future applications, it is very
important to understand the effect of periodic driving in the
properties of quantum wires. It can be argued that the
localization behavior of this kind of systems is its most basic
property. It is the goal of this paper to give the basic
foundations for the extension of localization theory to
periodically driven systems and to study the prominent features of
a simple model that includes both disorder and periodic driving.

In the absence of disorder, and for a 1-D tight binding system
with a band width $\Delta$, it was shown that, for
high-frequencies, the driven system is equivalent to a non-driven
system with a renormalized
band-width\cite{Holthaus92a,Holthaus92b},
\begin{equation}
\label{eq:delta} \Delta \rightarrow \Delta_{eff}= \Delta~
J_{0}\left(\frac{ev_{ac}d}{\hbar\omega}\right),
\end{equation}
where $v_{ac}$ is the amplitude of the ac-driving field and $d$ is
the spatial period of the lattice. At the zeros of this Bessel
function the width of the band goes to zero, the group velocity of
an electron wave-packet becomes zero and the electron becomes
effectively localized (however, the Floquet eigenstates continue
to be extended). This strong localization of a particle due to the
effects of a periodic driving has been called dynamical
localization\cite{Dunlap} or coherent destruction of
tunneling\cite{Grossmann91}.

The effect of a harmonic driving in the localization of a
disordered one-dimensional system was first studied by Holthaus
{\em et al.} \cite{Holthaus,HolthausPhilo}, who showed that a
driving field can have strong effects in the localization of a
particle moving in a disordered potential. When disorder is
introduced it is well-known that all the states in the
one-dimensional lattice are localized, with their degree of
localization controlled by the ratio between the bandwidth
$\Delta$ and the strength of the disorder $W$. Holthaus {\em et
al.} have argued that in the high-frequency limit the Floquet
states themselves become localized (in contrast with dynamical
localization) and their localization should depend on the ratio
$W/\Delta_{eff}$.
 
In a recent work \cite{MartinezMolina} we were able to generalize the
definition of localization length using a Floquet-Green function
formalism. Through a numerical calculation of the Floquet-Green
function for a one-dimensional tight-binding model with diagonal
disorder plus an external dipolar driving field we could confirm
quantitatively that in the high-frequency regime the effect of
the driving could be seen as a renormalization of the band-width.
More interestingly, in that work it was also shown that low-frequency driving can reduce the
localization of the electronic Floquet-states.

It is the purpose of this paper to present, in a detailed and self-contained manner, the
theory first used in \cite{MartinezMolina}. We also present new
 analytical calculations for the Floquet-Green function
in the high-frequency regime and explore
additional features of our model regarding the periodic dependence
of the localization length on energy as well as the effect of
inelastic processes in the delocalizing effect of low-frequency
driving fields.

Section \ref{sec:Floquet-Green} is devoted to the Floquet-Green
operator formalism which is the basis of all our results. Here 
we show its relationship to the transport properties of driven
systems. The generalization of the definition of localization
length is discussed in section \ref{sec:lambdadef}. In section
\ref{sec:model} we introduce the Hamiltonian and the method that
we follow to obtain the localization length for this system. In
section \ref{sec:frequencyamplitude} we examine the behavior of the localization length for different amplitudes of
the driving field and in different frequency regimes. In
particular, the results for $\lambda$ in the case of
high-frequency driving will be studied in subsection
\ref{sec:highw}. In this section we also derive some new analytical results for driven non-disordered systems which explain some features of the localization length of a disordered system in the high-frequency limit.   
Some previous obtained results for low-frequency delocalization will be presented
in subsection \ref{sec:loww}. New results for $\lambda$ as a
function of energy will be shown in section \ref{sec:band}. 
The previously unexplored effect of an effective inelastic scattering rate of the electrons
on the delocalization properties of low-frequency driving will be
studied in subsection \ref{sec:tau}. Finally, in section
\ref{sec:conclusions} we give some concluding remarks and
perspectives.

\section{Floquet-Green operator for time-periodic systems}
\label{sec:Floquet-Green}

The pioneering work of Shirley \cite{Shirley}, Zel'dovich
\cite{Zeldovich} and Sambe \cite{Sambe} laid down the theoretical
foundations for a complete treatment of time-periodic potentials,
based on the same mathematical tools already developed for
time-independent potentials. Of significant importance among these
tools is the Green function. A Floquet-Green function method for
the solution of radiative electron scattering in a strong laser
field was introduced by Faisal\cite{Faisal}. The definition and
application of the Green function formalism that fully exploits
the periodic time-dependence of the Hamiltonian has not been done
until recently.  In this section we show the complete
Floquet-Green operator formalism for general time-periodic
Hamiltonians, in the way that was introduced by one of the authors
in previous works \cite{Martinez03,Martinez05}.

We start by considering a general Hamiltonian of the form:

\begin{equation}
H(t)=H_0 + 2 V \textrm{cos}(\w t), \label{Hamiltonian1}
\end{equation}
were $H_0$ and $V$ are Hermitian operators in the Hilbert space
($\mathcal{H}$) of the system. Because of the periodicity of the
Hamiltonian, according to Floquet's theorem, the solutions to
Schrodinger's equation $i\hbar\frac{\partial}{\partial t}|\Psi
(t)\bra=H(x,t)|\Psi(t)\bra$ are of the form
\begin{equation}
|\Psi^e(t)\bra=e^{-i e t/\hbar} |\phi^e(t)\bra, \label{Psiphi}
\end{equation}
where $|\phi^e(t)\bra=|\phi^e(t+\frac{2\pi}{\omega})\bra$.

Inserting this into Schrodinger's equation one arrives at the
eigenvalue equation
\begin{equation}
H^{F}(t)|\phi^e(t)\bra= e |\phi^e(t)\bra , \label{Feigenval}
\end{equation}
where $H^{F}(t)$ is defined as
\begin{equation}
H^{F}(t)\equiv H(t)-i\hbar\frac{\partial}{\partial
t}.\label{floquetHam1}
\end{equation}

As pointed out by Sambe \cite{Sambe}, since Eq.(\ref{Feigenval})
is an eigenvalue equation, it can be solved using the standard
techniques developed for time-independent Hamiltonians, provided
we extend the Hilbert space to include the space of time-periodic
functions. In this extended space, the time parameter can be
treated as another degree of freedom of the system. A similar
concept is used in classical mechanics, and gives rise to the
concept of a "half" degree of freedom when dealing with
time-dependent Hamiltonians.

A suitable basis for this extended Hilbert space ($\mathcal{R}$)
is $\{ |\alpha\bra\otimes |n\bra,...\}$, where
$\{|\alpha\bra,..\}$ is a basis for the Hilbert space
$\mathcal{H}$ of the system, and we define $\ket t|n\bra=e^{-in\w
t}$, with $n$ integer. Clearly $\{|n\bra ,..\}$ spans the vector
space ($\mathcal{T}$ of periodic functions, and therefore,
$\mathcal{R=H\bigotimes T}$. In this basis, Eq.(\ref{Feigenval})
becomes a matrix eigenvalue equation of infinite dimension with an
infinite number of eigenvalues. It is not difficult to prove that
if $e_{\alpha,p}$ is an eigenvalue with corresponding eigenvector
$|\phi ^{e_{\alpha,p}}(t)\bra$, then $e_{\alpha,p} +m\hbar\omega$
is also an eigenvalue with corresponding eigenvector
$|\phi^{e_{\alpha,p} +m\hbar\omega}(t)\bra$= $e^{im\w
t}|\phi^{e_{\alpha,p}}(t)\bra$. Accordingly, the eigenstate
corresponding to the eigenvalue $e_{\alpha} + m\hbar\omega$ has
the same structure as the eigenstate corresponding to
$e_{\alpha,p}$, except that it is displaced by $m\hbar\omega$ on
the energy axis. Because of this, to find all the eigenvectors and
eigenvalues of the Floquet Hamiltonian one needs only to consider
$-\frac{1}{2}\hbar\omega\leq e <\frac{1}{2}\hbar\omega$. We will
use the letter $\eps$ to refer to the Floquet eigenvalues
restricted to this interval and call them "quasi-energies".
Clearly, any Floquet eigenvalue $e_{\alpha,p}$ can be written as
$e_{\alpha,p}=\eps_\alpha +p\hbar\omega$ for some
$-\frac{1}{2}\hbar\omega\leq\eps_{\alpha} <\frac{1}{2}\hbar\omega$
and some integer $p$. It can be shown that, in general, there are
N distinct quasi-energies (except for accidental degeneracies) if
the Hilbert space $\mathcal{H}$ is N-dimensional.

This periodic structure in the eigenvalues does not mean that the
"sideband" eigenstates have no relevance; they are also valid
solutions of Eq. (\ref{Feigenval}) and are essential for
completeness in the extended Hilbert space $\mathcal{R}$
\cite{Dresse99}.

The Floquet-Green operator corresponding to Eq. (\ref{Feigenval}),
is defined by the equation (see \cite{Martinez03,Martinez05}),

\begin{equation}
[\openone E-H^{F}(t')]\G(E,t',t")=\openone\delta_{\tau}(t'-t'')~~,
\label{definitionG}
\end{equation}
where $\delta_{\tau}(t)$ is the $\tau$-periodic delta function
($\tau=\frac{2\pi}{\w}$) and $\openone$ is the identity operator
in $\mathcal{H}$, which we will omit from now on.

Notice that this equation differs from the usual time-dependent
Green function on two instances: The Hamiltonian $H(t)$ is
replaced by $H^F (t)$, as defined in Eq.(\ref{floquetHam1}), and
also $\delta(t)$ is replaced by $\delta_\tau (t)$. With this
definition, the properties derived from the periodicity of the
Hamiltonian and its eigenfunctions are built-in features of the
Floquet-Green function. In terms of the complete (infinite) set
$\left\{ |\phi^{\alpha,p}(t)\bra\right\}$  of eigenfunctions of
the Floquet-Hamiltonian (Eqs.\ref{Feigenval},\ref{floquetHam1}),
the solution for Eq.(\ref{definitionG}) can be written as

\begin{equation}
\G(E,t',t'')=\sum_{\alpha,p} \frac{|\phi^{\alpha,p}(t')\bra \ket
\phi^{\alpha,p}(t'')|}{E-e_{\alpha,p}}~.
\end{equation}

>From the previous discussion about the eigenvalues and
eigenfunctions of the Floquet Hamiltonian, we can write the
Floquet-Green operator entirely in terms of the eigenfunctions
$|\phi^{{\alpha,0}}(t)\bra$, which correspond to values
{$e_{\alpha,p}$} between $-\frac{1}{2}\hbar\omega$ and
$\frac{1}{2}\hbar\omega$ :
\begin{equation}
\G(E,t',t'')=\sum_\alpha \sum_p e^{ip\w
(t'-t'')}\frac{|\phi^{{\alpha,0} }(t')\bra\ket
\phi^{{\alpha,0}}(t'')|}{E-\eps_\alpha-p\hbar\omega}~~,
\end{equation}
where $\gamma=1,...N$ for $\mathcal{H}$ being N-dimensional, and
$p=-\infty,...,\infty$.

Operating on both sides of this equation with
$\frac{1}{\tau^2}\int_0 ^\tau \int_0 ^\tau e^{im\w t'}e^{-in\w
t''}dt''dt'$ we obtain
\begin{equation}
\G^{m,n} (E)=\sum_{\alpha,p}
\frac{1}{E-\eps_{\alpha}-p\hbar\omega}|\phi^{{\alpha,0}}_{m+p}
\bra\ket\phi^{{\alpha,0}}_{n+p}|,
\label{Gandeigenveccomp1}
\end{equation}
where
\begin{equation}
\G^{m,n}(E)=\frac{1}{\tau^2}\int_0 ^\tau \int_0 ^\tau e^{im\w
t'}e^{-in\w t''}\G(E,t',t'')dt''dt'~, $$and$$
|\phi^{\alpha,0}_{m}\bra=\frac{1}{\tau}\int_0 ^\tau e^{im\w
t'}|\phi^{\alpha,0}(t')\bra dt'.\label{Gmn}
\end{equation}

At this point, a technical comment is in order. Notice that the
parameter $E$ in the Floquet-Green function is $\textit{not}$ a
quasienergy since it is not restricted to the first Brillouin
zone. The Floquet-Green function has poles along the whole real
axis. When considering a closed system with a Hamiltonian periodic
in time, it is well-know that the energy is not a conserved
quantity and that the system state is specified by its
quasienergy. From this point of view, one might think that $E$
should be restricted to the first Brillouin zone. However,
consider the physical situation in which the particle comes from
outside the system and one is interested in the probability for
the different processes that can occur to this particle in the
course of its interaction with a potential. These quantities can
all be obtained from the Green function, and in the case of
time-periodicity, from the Floquet-Green function. In particular,
$\G^{m,n}(E)$ gives the probability amplitude for a process where
the incident particle with energy $E$ enters in (Floquet) channel
$n$, and leaves the system through channel $m$. The meaning of
these "channels" is clear from the following property of the
Floquet-Green function: From Eq.(\ref{Gmn}),
$$\G^{m,n}(E)=\G^{m-k,n-k}(E+k\hbar\omega).$$
This means that the energy of the incoming particle can be thought
of as determining the incoming and outgoing channels. In fact,
$\G^{m,n}(E)$ can be interpreted as giving the probability of a
particle that comes in with energy $E+n$, absorbs $m-n$ photons
and leaves with energy $E+m$. From this it is clear that to
describe an incident particle with energy $E$ one can either
decompose $E=\eps+n\hbar\omega$ and use the quasi-energy $\eps$
along with an incident channel $n$, or simply use $E$ as the
energy and assume that the incident channel is $n=0$.  In the
following, we adopt the last view which we believe is the most
natural one.

>From the previous discussion, one can see that the quantities
$\G^{k,0}(E)$ provide all the information of the driven system,
\begin{equation}
\G^{k,0}(E)=\sum_{\alpha,p}
\frac{1}{E-\eps_{\alpha}-p\hbar\omega}|\phi^{\alpha,0}_{k+p}\bra\ket\phi^{\alpha,0}_{p}|~.
\label{Gk0}
\end{equation}

It is easy to show that in the limit $V \rightarrow 0$,
$G^{0,0}(E)\rightarrow G(E)$, where $G(E)$ is the usual Green
function for the autonomous system. Also, in this limit,
$G^{k\ne0,0}(E)\rightarrow 0$ (see ref.\cite{Martinez05} for an
equation that gives $G^{k,0}(E)$ in terms of $G^{0,0}(E)$). The
Floquet-Green operator components $G^{k,0}(E)$ are important
because transport properties of driven systems have been
formulated in terms of these components, which play a role similar
to the Green operator in the Landauer formalism for
conduction\cite{Kohler}. More specifically, it was found that the
average current through a 1-D driven system (coherent regime) can
be expressed as
\begin{equation}\nonumber
\bar{I}=\frac{\textit{e}}{\textit{h}}\sum_{k=-\infty}^\infty\int
dE\left\{T_{\ell
r}^{(k)}(E)\textit{f}_{r}(E)-T_{r \ell}^{(k)}
(E)\textit{f}_\ell (E)\right\},
\end{equation} where
\begin{equation}
T_{\ell r}^{(k)}(E)=\Gamma_\ell
(E+k\hbar\omega)\Gamma_r(E)|G_{1L}^{k,0}(E)|^2,$$
$$ T_{r \ell}^{(k)}(E)=\Gamma_r
(E+k\hbar\omega)\Gamma_\ell
(E)|G_{L1}^{k,0}(E)|^2,\label{Kohler's formula}
\end{equation}
denote the transmission probabilities for electrons from the
right(left) lead respectively, with energy $E$ and final energy
$E+k\hbar\omega$, i.e. the probability of a scattering event under
the absorption(emission) of $|k|$ photons, if $k>0$($k<0$). This
expression for the average current converges to the well-known
Landauer-B\"uttiker formalism \cite{Landauer} in the limit when
the driving amplitude goes to zero.

\section{Definition of localization length for driven
disordered systems}
\label{sec:lambdadef}

The localization length for 1-D disordered non-driven systems has
been defined in terms of Green functions \cite{Kramer}, through
the well-known relation
\begin{equation}
\label{eq:deflambdastatic} \frac{1}{\lambda
(E)}=-\lim_{L\rightarrow \infty} \frac{1}{L}
\left<\ln\left|G_{1L}(E)\right|\right>.
\end{equation}
This definition makes use of the proven fact that in a disordered
1-D potential, the wave functions for any energy decay
asymptotically in an exponential way with distance: $\Psi^E
(x)\approx e^{-x/\lambda(E)}$.

Using Eq.(\ref{eq:deflambdastatic}), and for a tight-binding
Anderson Hamiltonian (band-width $\Delta$ and diagonal disorder on
the on-site energies uniformly distributed in the interval
$[-W/2,W/2]$), the localization length is known to behave as
\cite{matematico}
\begin{equation}
\label{eq:resultl} \lambda=6.56 \left(\frac{\Delta}{W}\right)^2,
\end{equation}
valid for $\Delta/W$ not too large, and for an energy in the
middle of the band ($E=0$). Also, it is known that in autonomous
systems, $\lambda$ as a function of the energy $E$ follows an
inverse parabolic law with a maximum at $E=0$. This is so unless
$|E|>(\Delta+W)/2$ where the density of states is null.

Clearly, in the presence of a periodic driving, the definition in
Eq. (\ref{eq:deflambdastatic}) must be modified. Since the
Floquet-Green operator is the natural extension for time-periodic
systems of the Green operator for autonomous systems, and in view
of the expressions for the transport properties of driven
1D-systems, Eq.(\ref{Kohler's formula}), we consider the following
quantities as possible extensions of the concept of localization
length for driven systems:
\begin{equation}
\label{eq:deflambdafloquet}
\frac{1}{\lambda^{k}(E)}=-\lim_{L\rightarrow \infty} \frac{1}{L}
\left<\ln\left|G^{k,0}_{1L}(E)\right|\right>,
\end{equation}
in close analogy to Eq.(\ref{eq:deflambdastatic}). The different
quantities $G^{k,0}_{1L}(E)$ are associated with the probability
of a process where an electron starts with an energy $E$ at site
$1$ and ends at site $L$ with energy $E+k\hbar\omega$. In
principle, there could be a different localization length
associated with each one of these processes. However, since the
asymptotic behavior of the Floquet eigenstates is exponentially
decreasing (as we will show next), one can see that in the limit
$L\rightarrow \infty$, the dominant term in the sum over $p$ in
Eq. (\ref{Gk0}) always appears ($p$ can be positive or negative)
for any value of $k$. This implies that all the quantities
$G^{k,0}_{1L}(E)$ decay at the same rate with $L$ (even though
their values can be very different, depending on $V$ and
$\omega$.) Consequently, the quantities $\lambda^{(k)}(E)$ are all
identical for $V\neq 0$. However, as mentioned before, in the
limit $V \rightarrow 0$, $G^{0,0}(E)\rightarrow G(E)$  and
$G^{(k\ne0,0)}(E)\rightarrow 0$, where $G(E)$ is the Green
function of the time-independent system. This implies that the
best choice for defining the localization length of a driven
system is $\lambda^{(0)}(E)$, since it converges to the
localization length of the autonomous system when $V\rightarrow
0$. Due to these considerations, we finally define the
localization length for a driven system as:

\begin{equation}
\frac{1}{\lambda(E)}\equiv -\lim_{L\rightarrow \infty} \frac{1}{L}
\left<\ln\left|G^{0,0}_{1L}(E)\right|\right>.
\end{equation}

This definition of  localization length seems the most straight
forward and natural way to generalize this concept for a driven
system. It is the connection between our localization length and
the transport properties of a driven system that makes our
definition interesting and useful, with clear applications and
consequences in possible experimental setups.


\section{Model and method}
\label{sec:model}

In order to investigate the effect of a driven potential in the
localization properties of one-dimensional disordered systems we
have chosen the Anderson Hamiltonian with diagonal disorder. The
on-site energies $\epsilon_j$ are distributed uniformly in the
interval $[-W/2,W/2]$, where $W$ measures the strength of the
disorder. The driving potential is due to the presence of a
time-periodic spatially uniform field, e.g. the interaction
between an electron and an EM wave (dipolar approximation)
incident perpendicularly to the lattice, with its electric field
polarized along the lattice direction. (From now on we will use a
system of units in which $\hbar=1$, therefore, $\omega$ will refer
to the driving frequency measured in energy units.) Accordingly,
\begin{eqnarray}
\label{eq:Hamiltonian} H= &-\frac{\Delta}{4}\sum_j
\left(\left|j+1\right>\left<j\right|+
\left|j\right>\left<j\right|\right)+ \nonumber \\
& \sum_j \epsilon_j \left|j\right> \left<j\right|+2v\cos{\omega
t}\sum_j \left|j\right>j\left<j\right|.
\end{eqnarray}

In this work we will always use $\Delta=4$. For the calculation of
$G^{0,0}(E)$ we use a method developed by one of the authors (see
\cite{Martinez03} for details). For a periodic Hamiltonian of the
form $H(t)=H_0+2V\cos(\omega t)$, where $H_0$ and $V$ are any
time-independent operators in the Hilbert space of the system, the
Floquet-Green operator components satisfy
\begin{equation}
(E+k \omega-H_0)G^{k,0}-V(G^{k+1,0}+G^{k-1,0})=\delta_{k,0}.
\end{equation}
These equations can be solved using matrix continued fractions.
For the case $k=0$, one gets
\begin{equation}\label{Goo}
G^{0,0}(E)=(E-H_0-V_{eff}(E))^{-1},
\end{equation}
where
\begin{equation}\nonumber
V_{eff}=V^+_{eff}(E)+V^-_{eff}(E),
\end{equation}
with \begin{small}
\begin{equation} V_{{eff}} ^{\pm}(E)= V
\frac{1}{\displaystyle E \pm 1 \omega-H_0 -V
\frac{1}{\displaystyle E\pm 2 \omega-H_0
-V\frac{1}{~~\vdots~~}V}V}V~~. \label{eq:Veff}
\end{equation}
\end{small}

The convergence of equation (\ref{eq:Veff}) is system specific.
For our Hamiltonian, Eq.(1), the number of bands necessary to
ensure convergence increases linearly with $vL/\omega$. The
numerical performance of our method is determined by the speed in
the calculation of an $L \times L$-matrix inverse for each Floquet
sideband.

\begin{figure}
\begin{center}
\includegraphics[height=9cm,width=6cm,angle=-90]{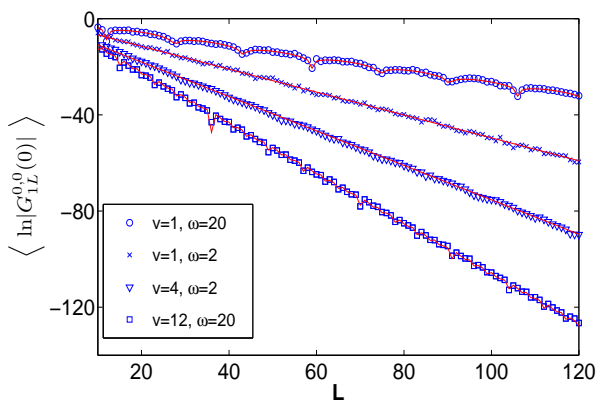}
\caption{\label{fig:scaling} $\left<\ln
G_{1L}^{0,0}(0)\right>$ as a function of the length of the system
$L$ for different values of disorder $W$, frequency $\omega$, and
field amplitude $v$.
High-frequency results, $\omega=20$, show regular oscillations on
top of the exponential decay, according to the equation
$G_{1L}^{0,0}(0)= AJ_{0}(2vL/\omega)\exp^{-L/\lambda}$. Low
frequency curves do not show these oscillations. The value of
$\lambda$ is the negative inverse slope of the linear fits. The
ensemble average was made with 1000 different realizations of
disorder.}
\end{center}
\end{figure}

To obtain the localization length from the Floquet-Green function,
we plot the ensemble average of $\ln{G_{1L}^{0,0}(0)}$ as a
function of the length of the system $L$. In Fig.\ref{fig:scaling}
we show some examples of the results obtained. As expected, a
straight line fits the data very well. The wave-functions decay
exponentially with distance. The negative slope of these curves
corresponds to the inverse of the localization length when $L \gg
\lambda$.

\section{Frequency and amplitude dependence of the localization
length}\label{sec:frequencyamplitude}
\subsection{High-frequency localization}
\label{sec:highw}

We now show results for the high frequency regime. Some results
for $\lambda(E=0)$ in this regime were already reported in
\cite{MartinezMolina}. The high frequency limit can be
characterized as the regime in which the absorption or emission of
any number of photons would leave the particle with an energy
outside of the region where the eigenenergies of the non-driven
system concentrate. This region is well known to have a width
$\Delta+ W$, and therefore, at the center of the band this
condition is satisfied when $\omega
> (\Delta + W)/2$.

We will first look at the high-frequency regime \textit{without}
disorder. For this case one can derive an analytical expression
for the Green function of the driven system, following the work of
M. Holthaus and D.W. Hone \cite{HolthausPhilo}. We will use this
expression to infer, for this frequency regime, the form of the
localization length for the disordered case.

The Floquet eigenstates can be written in the form of Houston
states \cite{Houston},
\begin{equation}
|\phi^{\kappa,0}(t)\bra=\sum_l |l\bra \ex \left ( iq_k (t)l d
-i\int_0 ^t d\tau[E(q_k (\tau))-\varepsilon(\kappa)]
\right)\label{Houston},
\end{equation}
where $q_k (t)=\kappa+2v \textrm{sin}\omega t$. Here $\kappa$ is
the quasimomentum of the Floquet state, which is connected to its
quasienergy through the dispersion relation
\begin{equation}
\varepsilon(\kappa)=J_0 (2v/\omega) E(\kappa) ~~mod~ \omega~~,
\end{equation}
and $E(\kappa)=-(\Delta/2) \textrm{cos}(\kappa)$.  The quantity
$E(q_k (\tau))$ can be written in terms of Bessel functions as
\begin{equation}
E(q_k (\tau))=-\frac{\Delta}{2}\sum_r J_r (v) \textrm{cos} (\kappa
+ r \omega \tau ).
\end{equation}

>From this analytical expressions for the Floquet eigenstates one
can construct the Floquet-Green operator, using Eq.(\ref{Gk0}) for
$k=0$. For that purpose we need to calculate the fourier
components of the Floquet eigenstates,

\begin{equation}
|\phi^{\kappa,0}_{m}\bra =\frac{1}{\tau}\int_0 ^\tau \ex (im\w
t')|\phi^{\kappa,0}(t')\bra dt'~~.
\end{equation}
>From Eq.(\ref{Houston}) we get,
\begin{equation}
\begin{split}
|\phi^{\kappa,0}_{m}\bra = &\sum_l |l\bra \frac{1}{\tau}\int_0 ^\tau
dt' \ex \{im\w t'+ iq_k (t')l d \\
 & -i\int_0 ^{t'} dt''\left[E(q_k
(t''))-\varepsilon(\kappa)\right] \}.
\end{split}
\end{equation}

At this point we can use the fact that in the high frequency
regime, $\Delta\ll \omega$ and therefore in this integral, to
lowest order in $\Delta/\omega$, we can ignore the contribution
from the terms that contain $E(q_k (\tau))$ and
$\varepsilon(\kappa)$ (both are proportional to $\Delta$). From
this the above expression reduces to

\begin{equation}
|\phi^{\kappa,0}_{m}\bra =\frac{1}{\tau}\sum_l |l\bra e^{ikld}
\int_0 ^\tau dt'e^{i \left[m\w t' +\frac{2vl}{\omega} sin(\omega
t') \right]}~,
\end{equation}
and using the identity
\begin{equation}
\ex (izsin\phi )= \sum_r \ex (ik\phi) J_k (z)~,
\end{equation}
one arrives at
\begin{equation}
|\phi^{\kappa,0}_{m}\bra =\sum_l e^{i\kappa l}|l\bra J_{-m}
(2vl/\omega)~.
\end{equation}
This expression is only valid at frequencies much higher than the
band-width $\Delta$. From this, and using Eq.(\ref{Gk0}), the
final expression for the quantity $G_{1L}^{0,0}$ for the system
\textit{without} disorder, in the high frequency regime is
\begin{equation}
G_{1L}^{0,0}(E)=J_0 (2v/\omega)~J_0
(2vL/\omega)\frac{1}{2\pi}\int_{-\pi}^{\pi} d \kappa
\frac{e^{-i\kappa(l-1)}} {E-\eps (\kappa)}. \label{HFG1L}
\end{equation}

\begin{figure}
\begin{center}
\includegraphics[width=8.5cm,height=6.5cm]{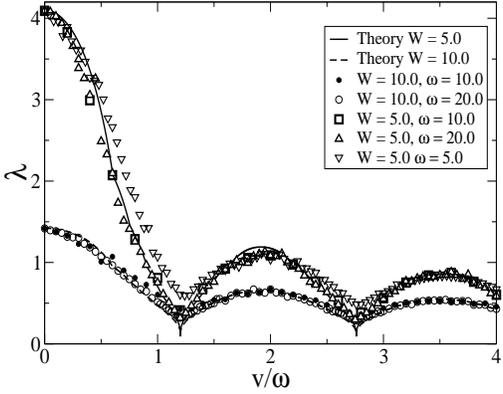}
\caption{ \label{fig:bessel} High frequency results. Results for
different values of disorder $W$ and $\omega$. The
corresponding results for a non-driven system with renormalized
bandwidth obtained from numerical data are shown with full lines
($W=5$) and dashed lines $W=10$.}
\end{center}
\end{figure}

As can be seen in Fig. \ref{fig:scaling}, for the second and last
sets of data ($\omega=20$) there are dips in the function
$\left<\ln(|G_{1L}^{0,0}(0)|)\right>$. The thin (red online)
continuous line represents the function $\ln (A J_0
(2v/\omega)\exp (-L/\lambda))$, which fits the high-frequency data
very well. The dips are clearly due to the zeros of the Bessel
function. These Bessel function factors, which appear in the
Floquet-Green function of the system \textit{without} disorder,
remain there after the disorder is introduced and the ensemble
averaging is taken. This is due to the fact that they do not
depend on any of the parameters of $H_0$.

>From Eq.(\ref{HFG1L}) above, we can also see that, apart from the
Bessel function factors, the net effect of the high-frequency
driving is to renormalize the band-width of the non-driven system,
i.e. instead of $E-E(\kappa)$ in the denominator, one gets
$E-\eps(\kappa)=E-J_0(2v/\omega)E(\kappa)$. This renormalization
in the bandwidth was shown by Holthaus and Hone to carry into the
localization length of a defect placed in the driven lattice
\cite{HolthausPhilo}. This means that $\lambda$ depends on the
ratio between the defect energy and the renormalized bandwidth.
>From this, one can expect that, for a driven disordered system in
the high-frequency regime, the results for $\lambda$ can be
obtained through
\begin{equation}\label{eq:lambda-bessel}
\lambda(v,\Delta)=\lambda\left(0,\Delta J_0 (2v/\omega)\right).
\end{equation}

As Fig. \ref{fig:bessel} shows, for different values of $\omega$
and for disorder $W=5$ and $W=10$, the numerical data is in
excellent agreement with this hypothesis. The minima of $\lambda$
correspond to the zeros of the Bessel function $J_0(2v/\omega)$.
The slight deviations are more significant for smaller values of
$\omega$, possibly due to next order corrections in $v/\omega$,
since we expect Eq.(\ref{eq:lambda-bessel}) to be strictly valid
only for infinite frequency.


\subsection{Low-frequency delocalization}
\label{sec:loww}

As it was already found in \cite{MartinezMolina}, the behavior of $\lambda$
for low-frequency driving is very different from the
high-frequency case. As it can be seen in Fig.\ref{fig:g05} a), $\lambda$ as a function of $v/\omega$
initially \textit{increases}. This delocalization can only occur
for $\omega < (\Delta+W)/2$. As a function of $v$, the
localization length in the low-frequency regime increases, then
reaches a maximum and finally decreases. This maximum value
increases as the frequency is decreased.


In \cite{MartinezMolina} an intuitive interpretation of this
result was proposed. The driving allows the electron to exchange
energy with the external field and to explore new regions of the
phase space. For an electron of initial energy $E_0$, turning on
an ac-field allows it to exchange energy with the field
(quantum-mechanically) through an integer number of photons. Due
to this, the electronic state becomes a superposition with
energies $E_n =E_0 \pm n \omega$. Each electron path corresponding
to each one of these energies behaves as an effective channel with
a relative weight that depends on the value of the field amplitude
$v$. If the first of these new channels is outside the band-width
it will have a negligible probability of having a localization
length higher than the original one without driving. However, if
this new channel is inside the band-width of the disordered wire
it will have a finite probability of having a higher localization
length than the original path of the electron with energy $E_0$.
As the total localization length of the system will depend on the
less localized of the electron paths, the ensemble-averaged
localization length will increase with respect to the autonomous
case. Given that the number of effective channels that fit into an
energy region of size $\Delta+W$ is inversely proportional to the
frequency, the localization length will therefore increase with
decreasing frequency as it was found in \cite{MartinezMolina}.

These findings can have an important impact on the transport
properties of disordered nanowires in the presence of electric
ac-fields. In principle, at high-frequency, and due to dynamical
localization, the ac-field will decrease the electric current
through the device, whereas, at low-frequency, an ac-field can be
used to increase the localization length so that it becomes
greater than the length of the wire, thereby producing an increase
of several orders of magnitude in the current that passes through
the nanowire.

\section{Energy dependence of the localization length}
\label{sec:band}

\begin{figure}
\begin{center}
\includegraphics[width=8.5cm,height=6.5cm]{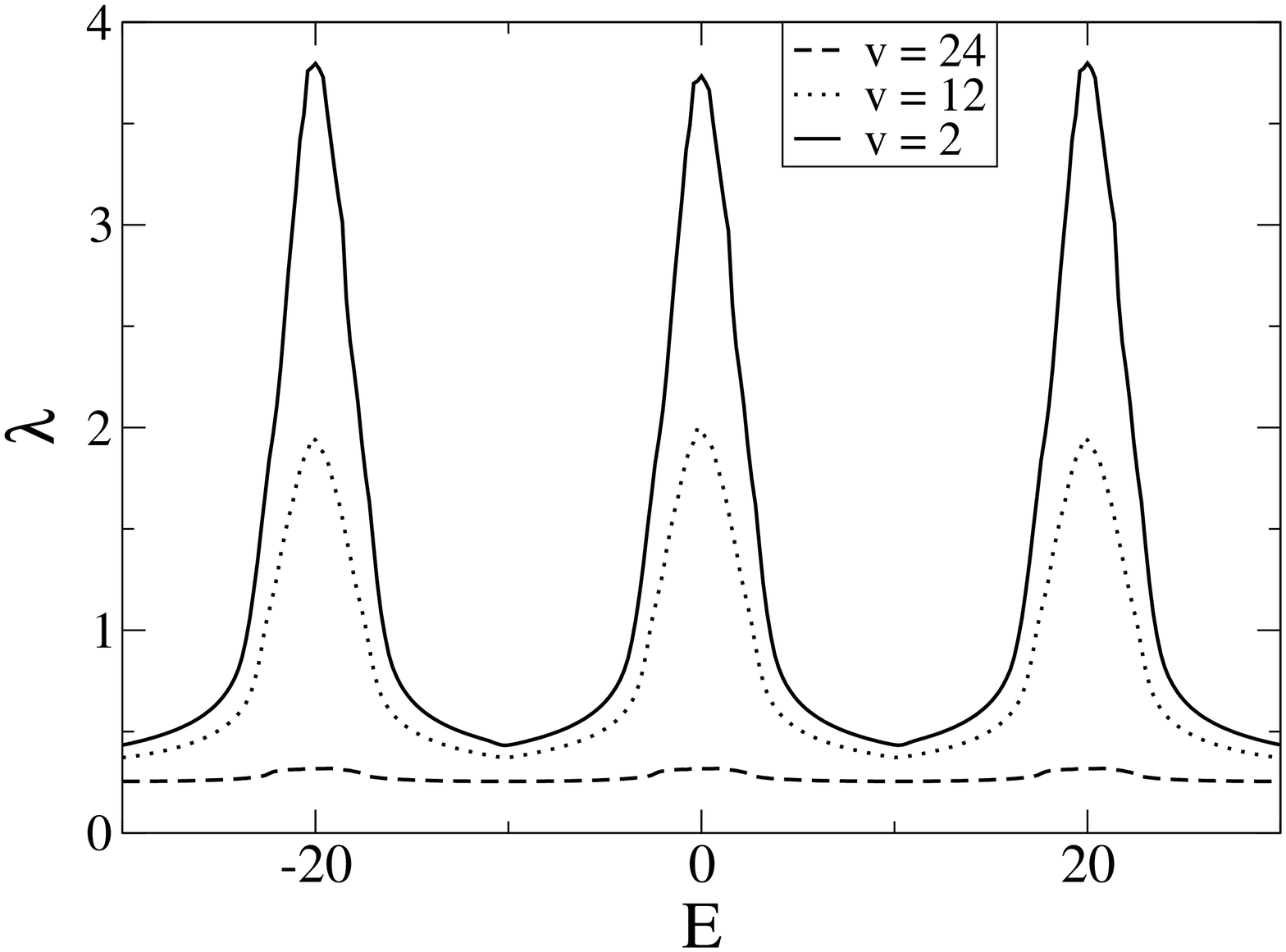}
\caption{\label{fig:banda1} Localization length $\lambda$ as a
function of the energy for $W=5$, and $\omega=20$.}
\end{center}
\end{figure}


As we have mentioned previously, $\lambda$ as a function of the
energy for the autonomous system follows an inverse parabolic law
with a maximum at the band-center $E=0$. The aim of this section
is to study how this dependence is modified in the presence of
driving. First of all we will show that due to the structure of
the Floquet-Green function the localization length $\lambda(E)$ is
a periodic function of the energy with period $\omega$, as it can
be seen in Fig.(\ref{fig:banda1}). It is easy to show rigourously
that for any $V\neq 0$, the localization length is periodic on the
energy. Before proving this, we emphasize that this is not a
trivial statement. This periodicity is \textit{not} present in the
Floquet-Green operator from which the localization length is
obtained. This can be shown as follows:

>From Eq.(\ref{Goo}) we have
\begin{equation}
G^{0,0}(E+ \omega)=(E-H_0-V_{eff}^{+}(E+ \omega)- V_{eff}^{-}(E+
\omega))^{-1},\label{GooE+1}
\end{equation}
and from Eq.(\ref{eq:Veff}),
\begin{equation}
V_{eff}^{+}(E+ \omega) =E+ \omega - H_0 -
V~[V_{eff}^{+}(E)]^{-1}~V,
\end{equation}
\begin{equation}
V_{eff}^{-}(E+ \omega) =V~\frac{1}{E - H_0-V_{eff}^{-1}(E)}~V.
\end{equation}
Using this in Eq.(\ref{GooE+1}), one gets
\begin{equation}
\begin{aligned}
G^{0,0}(E+ \omega) &=
V^{-1}V_{eff}^{+}(E)G^{0,0}(E)V_{eff}^{+}(E)V^{-1}
\\&+ V^{-1} V_{eff}^{+}(E)V^{-1}.
\end{aligned}
\end{equation}
>From this equation it is clear that in general $G^{0,0}(E+ \omega)\neq
G^{0,0}(E)$.

We now proceed to show that, despite of the previous
considerations, the localization length is indeed periodic,
$\lambda(E)=\lambda(E+\omega)$. As we have commented before, from
Eq.(\ref{Gmn}) it is easy to see that
$G^{m,n}(E)=G^{m-n,0}(E+n\omega)$, and in particular,
\begin{equation}
G^{0,-1}(E+ \omega)=G^{1,0}(E).\label{GdiffE}
\end{equation}
To continue our proof, we consider the quantities
$\lambda^{k,j}(E)$ defined by Eq.(\ref{eq:deflambdafloquet}), but
with the upper index $0$ in the Floquet-Green function replaced by
$j$. From Eq.(\ref{Gandeigenveccomp1}) we can see that exchanging
the two upper indexes in $\lambda^{k,j}(E)$ we get the same value.
In particular, $\lambda^{k,0}(E)=\lambda^{0,k}(E)$. Also, as we
have shown before, all these different localization lengths are
equal to each other. Specifically,
\begin{equation}
\lambda^{0,0}(E)=\lambda^{1,0}(E)~,$$and$$
\lambda^{0,0}(E+\omega)=\lambda^{0,-1}(E+\omega).
\end{equation}

Using this, along with Eq.(\ref{GdiffE}) we arrive at
\begin{equation}
\lambda^{0,0}(E)=\lambda^{1,0}(E)=\lambda^{0,-1}(E+ \omega)=
\lambda^{0,0}(E+ \omega),
\end{equation}
as claimed.

\begin{figure}
\begin{center}
\includegraphics[width=8.5cm,height=6.5cm]{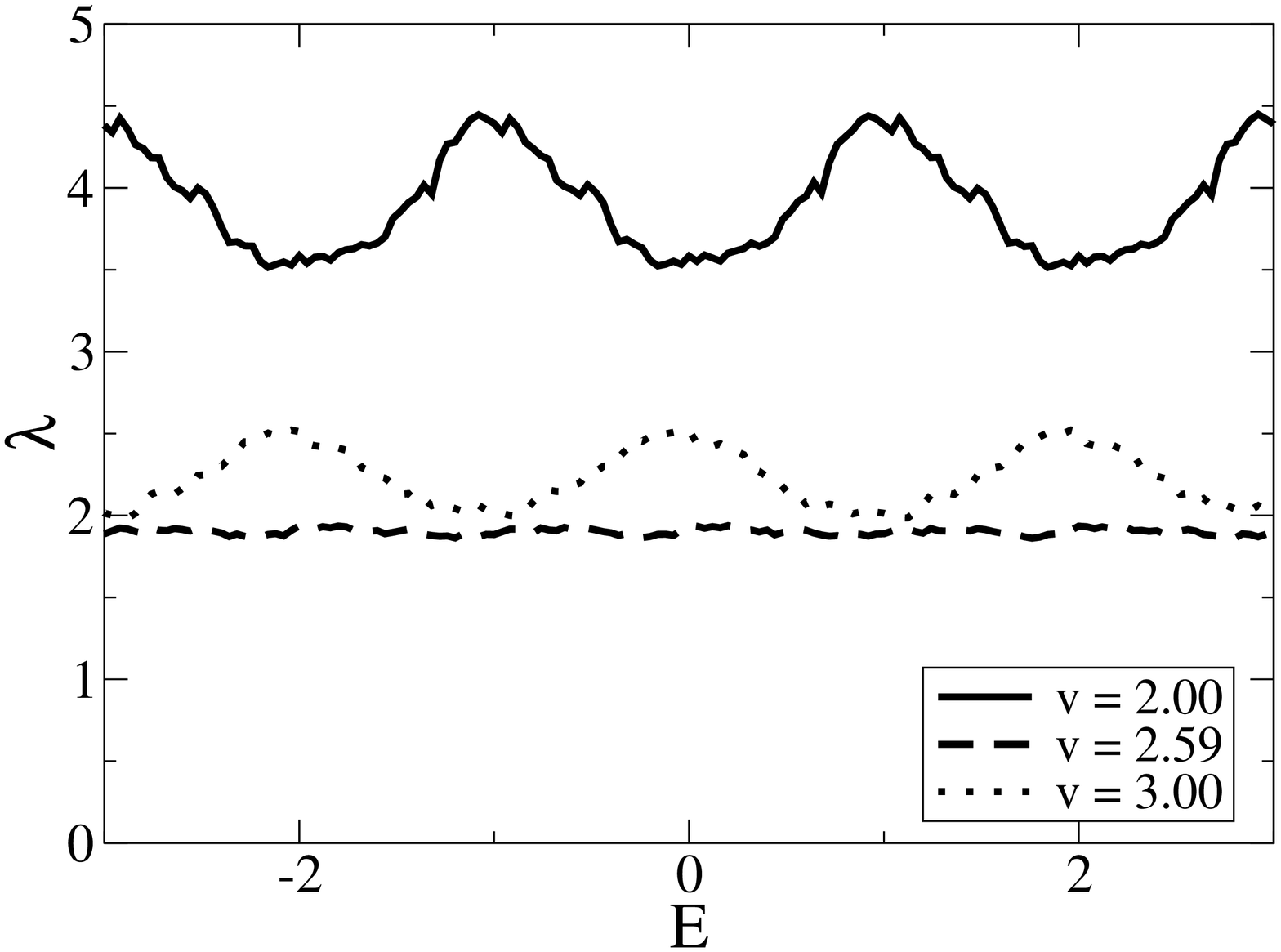}
\caption{\label{fig:banda2} Localization length $\lambda$ as a
function of the energy for $W=5$, frequency $\omega=2$ and three different
values of the field amplitude $v$.}
\end{center}
\end{figure}

In Fig. \ref{fig:banda1} we can see the clearly periodic behavior
of $\lambda$ as a function of the energy, for the high-frequency
regime. In this case, the form of the function $\lambda(E)$ is
similar to the form for the autonomous system. The only
differences are the previously explained renormalization of the
band width and the fact that this structure is repeated
periodically. In the low-frequency case the form of this function
changes. There is not always a maximum at the band center. In this
case one can often find a minimum at $E=0$ and two maxima located
symmetrically around this point (due to the symmetrical
distribution of disorder used), which are separated by a
difference in energy equal to the frequency of the driving field.
In Fig. \ref{fig:banda2} we can see how increasing the amplitude
$v$ we can go from the case of the two symmetric maxima at
$E=-\omega/2$ and $E=\omega/2$ to the band-center maximum at $E=0$
in a case with disorder $W=5$ and frequency $\omega=2$. There is
one intermediate critical value of $v$ where $\lambda(E)$ is
completely flat, $v=2.59$ in the figure. This value coincides with
the first local minima of $\lambda$ as a function of $v$. In
general we can say that in the very high-frequency regime there is
always a maximum for $\lambda$ at $E=0$ and that for very
low-frequency the localization length as a function of energy
changes very little, with small oscillations around an average
value. For values of the frequency in between these two extremes,
the behavior of $\lambda$ with energy depends on the specific
value of the different parameters of the system.


\section{Effect of an inelastic scattering rate and the adiabatic limit}
\label{sec:tau}

In real systems, inelastic effects will occur in the system and
the exchange of energy between the electrons and the periodic
external field will be hampered. In this section we want to
address the effect of an inelastic scattering rate which will
appear in physical systems due to scattering with phonons in the
crystal or by other mechanisms. The microscopic modelling of these
inelastic effects is out of the scope of this work. We introduce
here a phenomenological parameter, the inelastic scattering time
$\tau_{in}$. We are interested in studying the effects of this
inelastic scattering time in the localization properties of
disordered one-dimensional systems. We will show next that this
inelastic time creates a frequency cut-off for the delocalization
induced by low-frequency driving. We will also address the
relationship between these results and the adiabatic limit
$\omega\rightarrow 0$.


\begin{figure}
\begin{center}
\includegraphics[width=6.5cm,height=5.3cm]{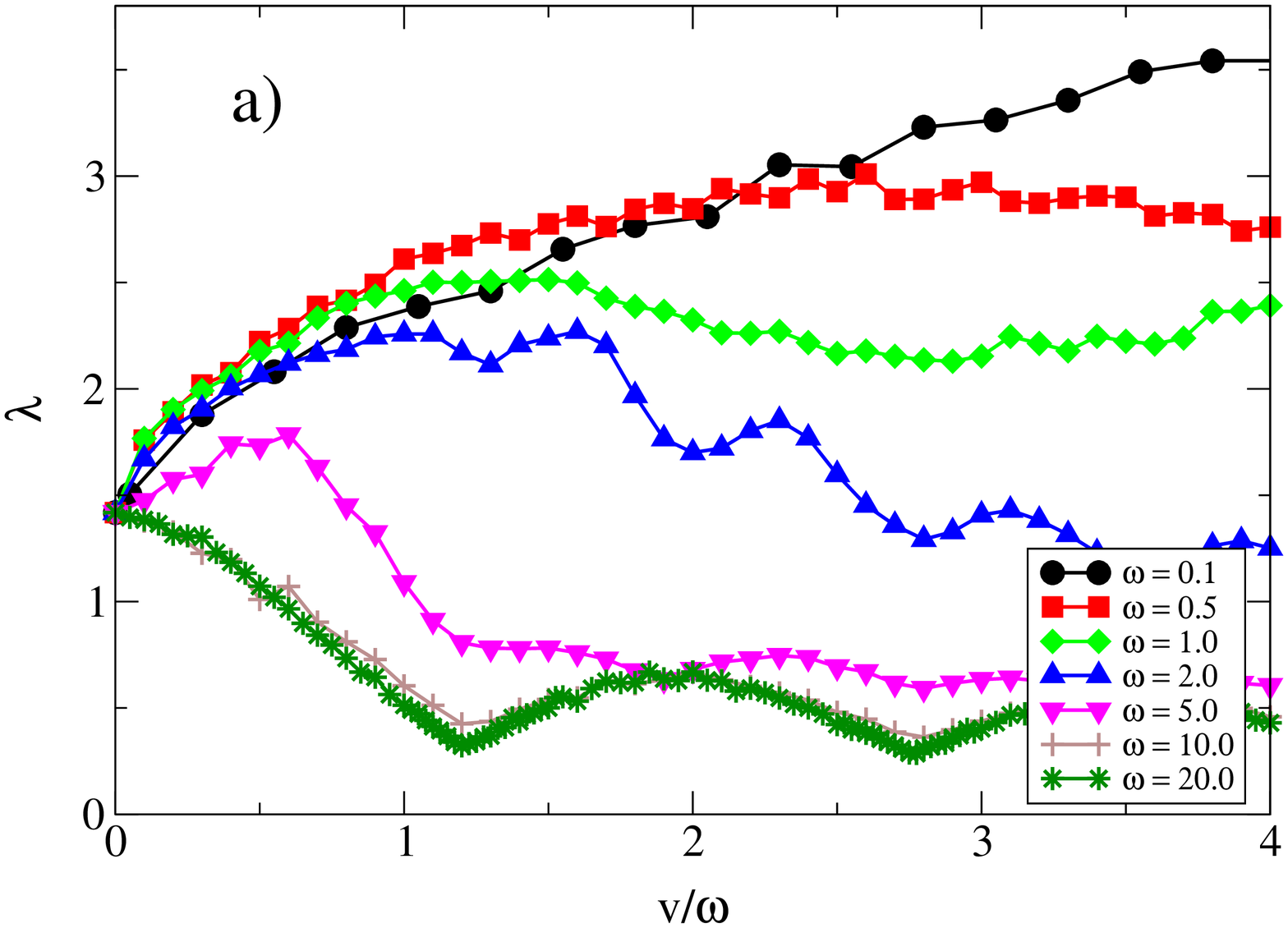}
\includegraphics[width=6.5cm,height=5.3cm]{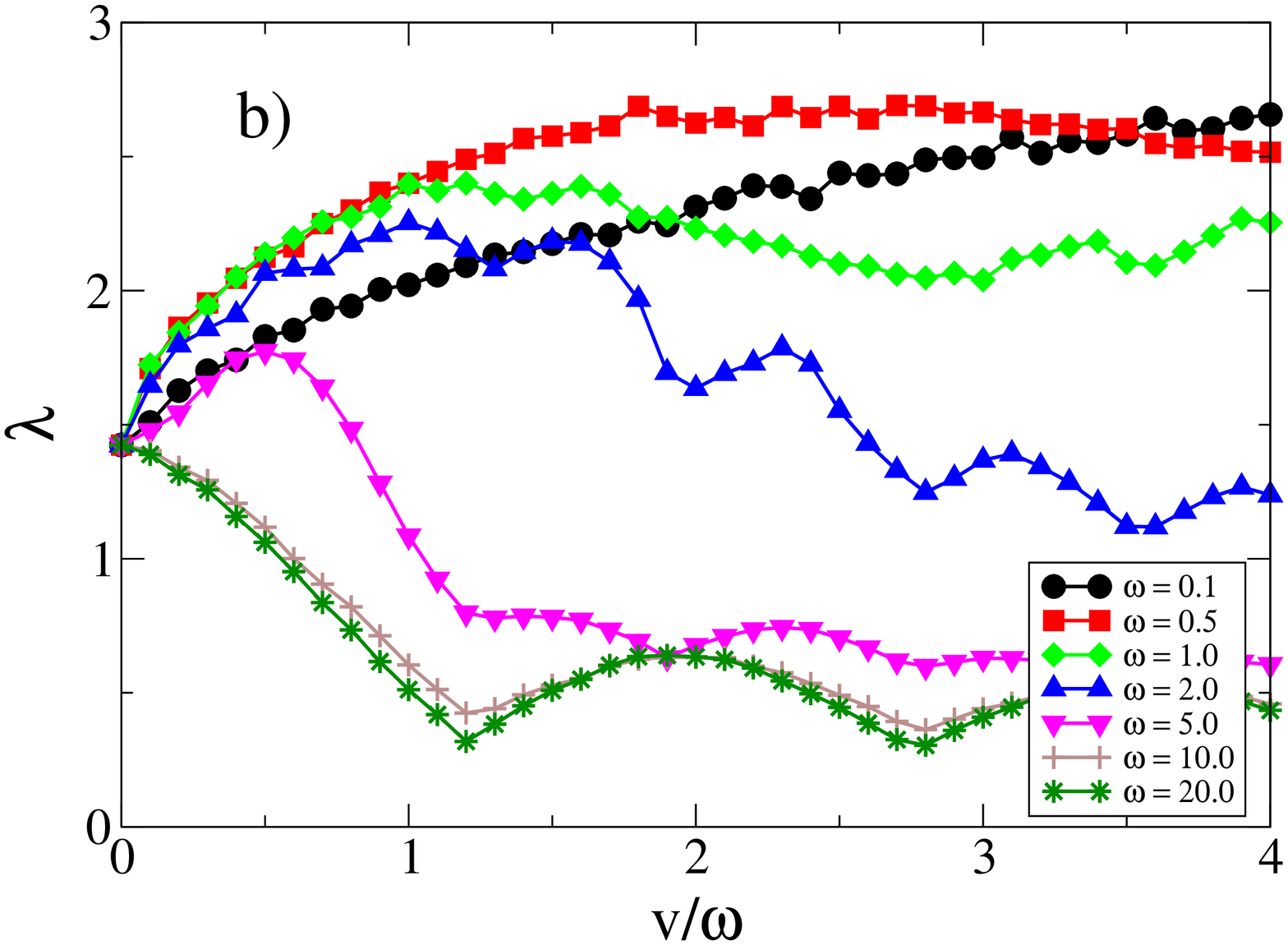}
\includegraphics[width=6.5cm,height=5.3cm]{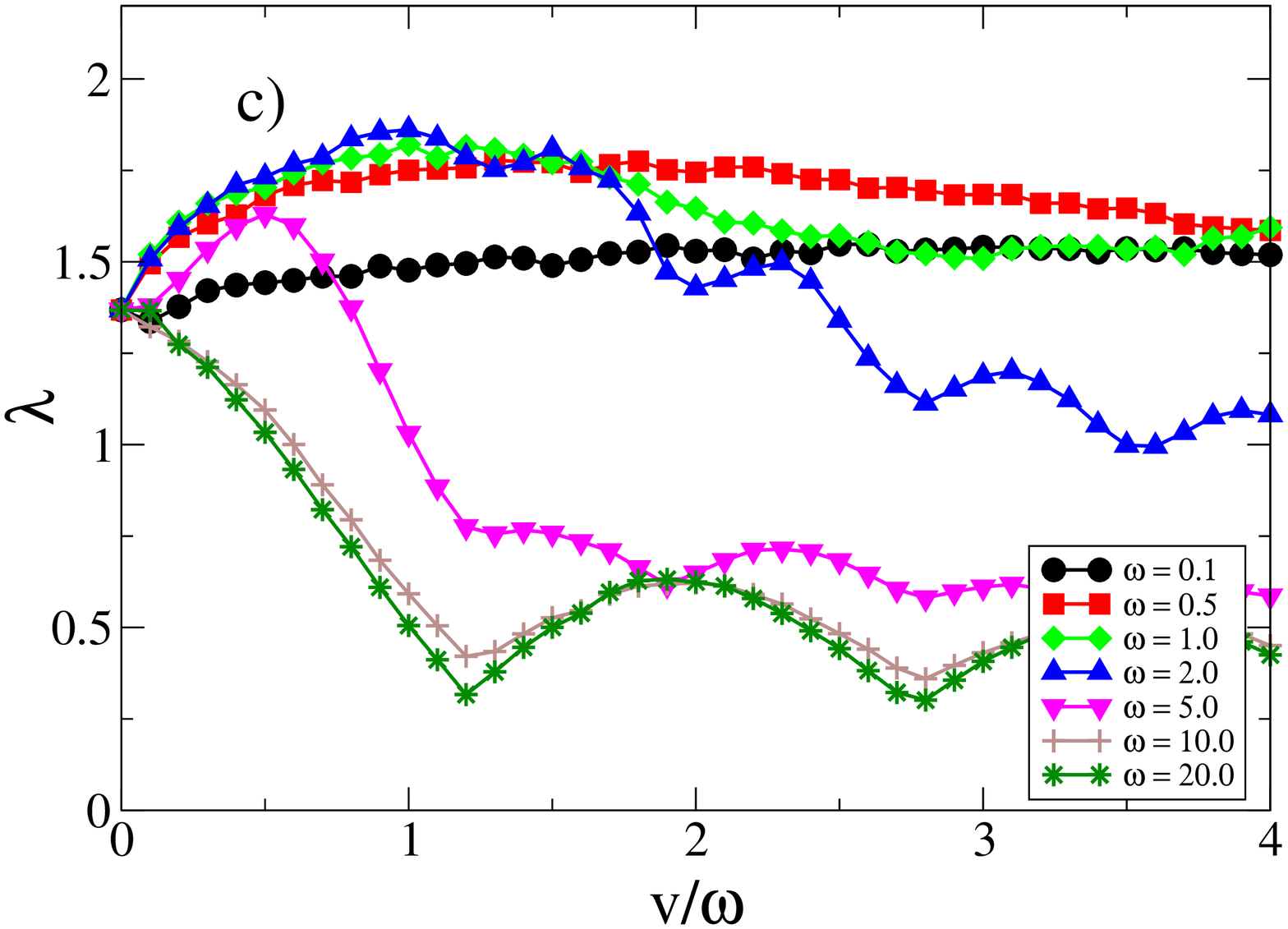}
\includegraphics[width=6.5cm,height=5.3cm]{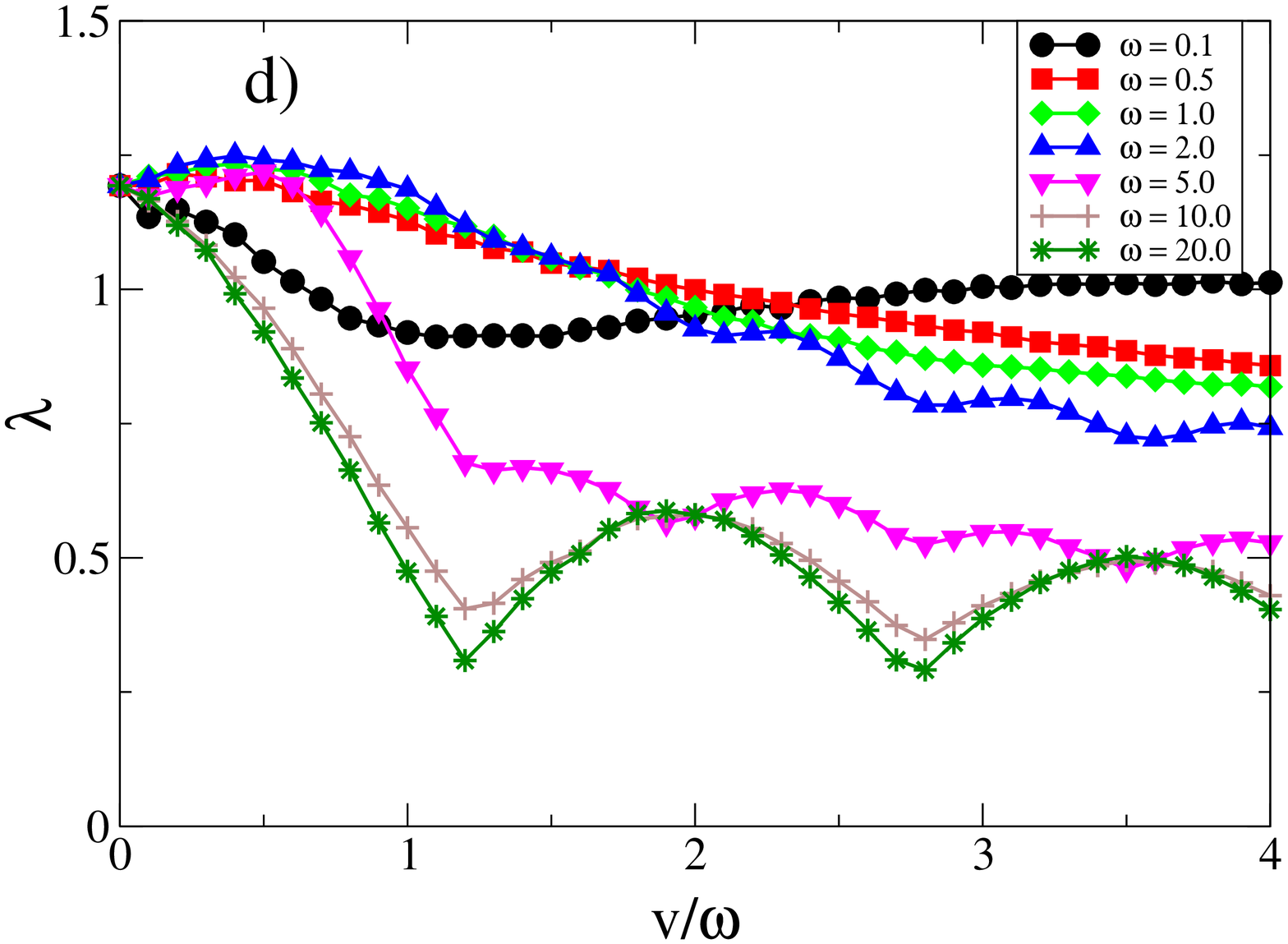}
\caption{\label{fig:g05} Effect of an inelastic scattering rate in
the localization length. $\lambda$ as a function of $v/\omega$ for
disorder $W=10$ and different values of the frequency $\omega$. In
a) $\Gamma=0$, b) $\Gamma=0.01$, c) $\Gamma=0.1$, d)$\Gamma=0.5$.}
\end{center}
\end{figure}

%


We introduce now a phenomenological inelastic scattering rate
$\Gamma=1/\tau_{in}$ just as an imaginary part of the energy
\cite{Mahan} in
the Floquet-Green function, Eqs. (\ref{Goo}) and (\ref{eq:Veff}).
We have studied four cases for disorder $W=10$. The localization
length as a function of $v/\omega$ is shown in Fig. \ref{fig:g05}
for a)$\Gamma=0$, b) $\Gamma=0.01$, c) $\Gamma=0.1$ and d)
$\Gamma=0.5$.

The value of $\lambda$ for $v=0$ is also reduced due to the
inelastic scattering rate $\Gamma$, from $\lambda=1.41$ when
$\Gamma=0.0$ to $\lambda=1.37$ when $\Gamma=0.1$, to
$\lambda=1.19$ when $\Gamma=0.5$.  For $\Gamma=0.01$ the reduction
could not be seen with the ensemble average performed. In the
high-frequency case the behavior of $\lambda$ as a function of $v$
is unchanged by the presence of inelastic scattering, apart from
this global reduction. We can clearly see the effect of dynamical
localization previously discussed and the shape of the Bessel
functions when $\omega$ is bigger than the band-width.

When we enter the low-frequency regime, the effect of the
inelastic scattering is to reduce the delocalization induced by
the low-frequency driving. The effect is completely suppressed for
frequencies $\omega < \Gamma$. See, for example, the case
$\omega=0.1$ when $\Gamma=0.5$. This can be easily understood by
considering that since the exchange of photons with the external
field allows the electron to explore more parts of the phase space
and to "choose" the best ones for propagation, then clearly,
energy dissipation due to inelastic processes will limit this
exploration therefore degrading the ability of the driving field
to delocalize the electron wave function.
\begin{center}
\begin{figure}
\includegraphics[width=8.5cm,height=6.5cm]{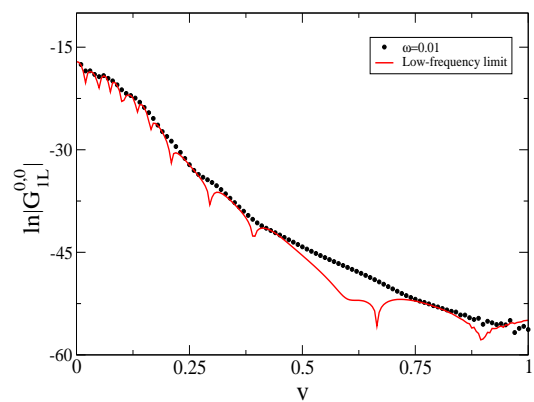}
\caption{\label{fig:low} Adiabatic limit for $\ln|G_{1L}|$ as a
function of $v$ for a particular realization of disorder with
$W=10$ and $\Gamma=0.5$. A comparison is made with the results for
$\omega=0.01$.}
\end{figure}
\end{center}

The inelastic scattering rate $\Gamma$ is also very important to
correctly obtain the static limit $\omega=0$ from our formalism.
In the adiabatic limit, when $\omega \rightarrow 0$, it has been
shown by Moskalets and B\"uttiker \cite{MoskaletsButtiker}, that
the Floquet scattering matrix agrees with the time average (over
one cycle) of the stationary scattering matrix. This property
implies that the Floquet-Green function $G^{0}(E+i\Gamma)$ in such
limit will converge to the time-average of the Green function of
the stationary system.
\begin{equation}
G^{(0)}(E+i\Gamma) = \frac{1}{T}\int_0^T G_0(v \cos \omega
t,E+i\Gamma) dt \,\,\,\, \textrm{when} \,\,\,\, \omega \rightarrow
0,
\end{equation}
where $G_0(v,E)$ is the Green function of the Anderson model with
a linear potential.

\begin{figure}
\begin{center}
\includegraphics[width=8cm,height=5cm]{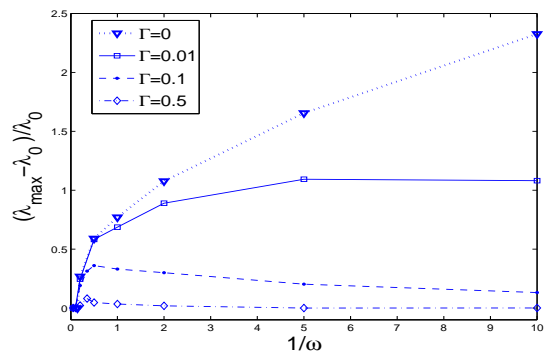}
\caption{\label{fig:eband} Behavior of $\lambda_{max}$ as a
function of $1/\omega$ for $W=10$ and different values of
the inelastic scattering rate $\Gamma$.}
\end{center}
\end{figure}

We need to include a small imaginary part to the energy to assure
the convergence of the time integral, since without it, the poles
of our Green function lie directly on the real energy axis. The
adiabatic regime, where the right-hand side of the above equation
is a good approximation of the left-hand side, is achieved when
$\omega<\Gamma$. This energy relaxation therefore becomes a
limiting time-scale for the delocalization effect of low-frequency
driving. From the numerical data in Fig. (\ref{fig:eband}), we see
that the monotonic increase of $\lambda$ with $1/\omega$ stops at
$\omega\gtrsim 20\Gamma$. When the condition $\omega<\Gamma$ is
achieved, the value of $\lambda$ is obtained from the time-average
of the Green function of the static system, which implies that in
this regime there can not be any delocalizing effect due to the
field since in the static limit the electric field does not
increase the localization length for any field amplitude
($\lambda_{max}=\lambda_0$). To obtain the correct static limit
using our formalism and for the purely theoretical case when there
is no inelastic scattering, one needs to take the limits
$\Gamma\rightarrow 0, \omega\rightarrow 0$, provided
$\omega<\Gamma$.

\section{Conclusions}
\label{sec:conclusions}

We have explored non-perturbatively, both for high and low
frequency, the localization of disordered one-dimensional
tight-binding lattices using a Floquet-Green function formalism
which makes use of matrix continued fractions\cite{Martinez03}.
This formalism allowed us to calculate directly the localization
length ($\lambda$) generalizing the standard definition for
autonomous systems. This quantity is very important in the study
of disordered systems and its properties are well known for
disordered non-driven systems. In a 1-D tight-binding model with
diagonal disorder we have found that in the high-frequency limit,
$\lambda$ is renormalized as a consequence of dynamical
localization. In this regime, the localization length always
\textit{decreases} with the amplitude of the driving field. Our
results show that the high-frequency regime in this model is
reached for $  \omega> (\Delta+W)/2$. For low-frequency, where $
\omega < (\Delta+W)/2$, we found the \textit{opposite} behavior:
$\lambda$ can be significatively \textit{increased} in the
presence of a driving field. According to this, one can say that
Anderson localization in this kind of disordered system is
enhanced by high-frequency driving and diminished by low-frequency
driving. For low frequency driving, each additional Floquet
channel created by the driving provides the electron with
additional paths. These will differ in their degree of
localization, with some having a smaller and others having a
greater localization length as compared to the non-driven case.
Since this quantity is given by the paths which have maximum
extension in space, we conclude that after the ensemble average
has been performed, additional propagation channels should always
contribute to increase the localization length. This simple
picture however, does not apply to the high-frequency case because
new paths introduced by the absorbtion or emission of one or more
photons always have localization lengths smaller than in the
non-driven case.

We have shown that due to the Floquet structure of the states, the
localization length in a driven system is periodic in energy, with
the periodicity given by the frequency of the driving field. For
frequencies well above $(\Delta+W)/2$, the maximum of $\lambda(E)$
is always at $E=0$ as it is in the autonomous system. For
frequencies well below this value, the localization length does
not change very much with energy. For intermediate values of
frequency, the localization length can have either a maximum or a
minimum at $E=0$, depending on the specific parameters of the
system.


We have explored the limit of very-low frequencies. We have shown
that the adiabatic limit for this theory is only well defined if
we add an imaginary part to the energy in the Floquet-Green
function. This quantity has the physical meaning of an energy
dissipation rate or inelastic scattering rate $\Gamma$ and has
important physical consequences. The dynamical delocalization
effect discussed in our work will be limited by $\Gamma$ and will
be completely suppressed for frequencies $\omega < \Gamma$.

The different results analyzed in our work should have important
experimental consequences in the field of coherent transport. They
lead to new avenues for the control of the localization and
transport properties of quantum wires and also for atoms in
optical traps, which have also been proposed as a testing ground
for Anderson localization. Recently, several experimental results
with BEC atoms in random potentials have been reported
\cite{coldatomsII}. For trapped cold atoms, a lattice potential
can be implemented by far detuned counter-propagating laser beams,
the AC driving can be obtained using a periodic phase-shift
between the beams \cite{coldatomsI} and the random potential can
be obtained with a superimposed random speckle pattern
\cite{speckles}.  We believe our results could lead to very
interesting effects in these systems.




\end{document}